\documentclass[3p,times]{elsarticle}

\usepackage{ecrc}

\usepackage{epstopdf}

\volume{00}

\firstpage{1}

\journalname{Nuclear Physics A}

\runauth{K. Fukushima}

\jid{nupha}

\jnltitlelogo{Nuclear Physics A}

\usepackage{graphicx}
\usepackage{amsmath,amssymb}
\newcommand{\LQCD}{\Lambda_{\text{QCD}}}
\newcommand{\Nc}{N_{\text{c}}}
\newcommand{\Tc}{T_{\text{c}}}
\newcommand{\muB}{\mu_\text{B}}
\newcommand{\muq}{\mu_\text{q}}
\newcommand{\bj}{\boldsymbol{j}}
\newcommand{\bE}{\boldsymbol{E}}
\newcommand{\bB}{\boldsymbol{B}}

\begin{document}

\begin{frontmatter}



\title{Baryonic matter and beyond}

%% Single author (and collaboration) - please insert
\author{Kenji Fukushima}
%\fntext[col1] {A list of members of the XYZ Collaboration and acknowledgements can be found at the end of this issue.}
\address{Department of Physics, The University of Tokyo,
         7-3-1 Hongo, Bunkyo-ku, Tokyo 113-0033, Japan}

%% For multiple authors, replace the above by:

%\author[label1]{Author1}
%\author[label2]{Author2}

%\address[label1]{Address 1}
%\address[label2]{Address 2}

\begin{abstract}
We summarize recent developments in identifying the ground state of
dense baryonic matter and beyond.  The topics include deconfinement
from baryonic matter to quark matter, a diquark mixture, topological
effect coupled with chirality and density, and inhomogeneous chiral
condensates.
\end{abstract}


\end{frontmatter}

% \linenumbers


\section{Onset of deconfinement at high temperature}

Color \textit{confinement} in QCD is still a big mystery and some
people may want to think that color \textit{deconfinement} would be
more understandable; otherwise, we have no idea what we are talking
about when it comes to the so-called quark-gluon plasma (QGP).  It is
our belief that the QGP has been created in the laboratory by means of
nucleus-nucleus collision at high enough energy at RHIC in Brookhaven
and LHC in CERN.\ \ It is not our present aim to challenge this widely
accepted interpretation.  The goal of this contribution should be to
encourage for physics opportunities of the lower-energy
nucleus-nucleus collision mainly in the context of the QCD phase
diagram research. Since our theoretical understanding in
finite-density QCD is so poor, what we can do the best is to make a
guess from what we have known.  In other words, we must carry
extrapolation out from known physical states, or preferably
interpolation between known states if possible.

\begin{figure}
 \begin{center}
 {\huge (a)} \hspace{0.38\textwidth} {\huge (b)}\\[1em]
 \includegraphics[width=0.4\textwidth]{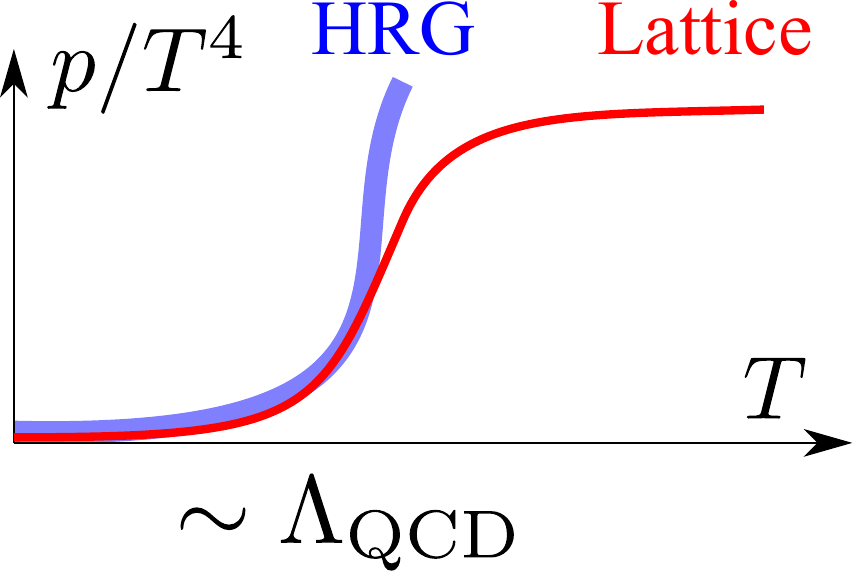} \hspace{2em}
 \includegraphics[width=0.4\textwidth]{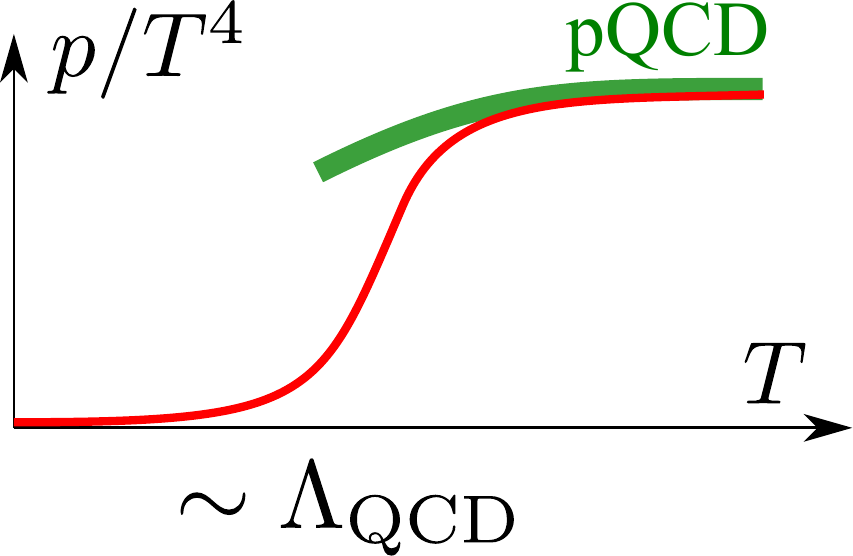}
 \end{center}
 \caption{Schematic figures of the normalized pressure as a function
   of $T$.}
 \label{fig:thermo}
\end{figure}

In this section we discuss the extrapolation from high temperature and
zero density along the phase transition.  Nowadays it is said that the
lattice QCD simulation has unveiled all features of QCD thermodynamics
as long as the baryon chemical potential is much smaller than the
temperature $T$.  Indeed the lattice QCD simulation is so powerful
that precise data of the pressure, the internal energy, the entropy
density, etc are available as functions of $T$.  Historically
speaking, because the QCD transition temperature turned out to be
$\Tc\sim\LQCD$ in early lattice QCD studies, the dream for the QGP
factory came to reality~\cite{Baym:2001in,Fukushima:2008pe}.
Figure~\ref{fig:thermo} is a schematic figure of the dimensionless
pressure $p/T^4$ that effectively measures the physical degrees of
freedom in the thermal system.  It is quite conceivable to interpret
the rapid rise in $p/T^4$ as the liberation of colored particles;
$\Nc$ quarks and $(\Nc^2-1)$ gluons.

Thus, answering the following question seems easy; where do you find
color deconfinement?  Your answer would be that it takes place at
$\Tc\sim\LQCD$ where $p/T^4$ starts growing up.  This conservative
answer is of course not wrong, but not completely satisfactory.  For
the increasing behavior of $p/T^4$ another interpretation is
possible, as illustrated by the HRG line in Fig.~\ref{fig:thermo}~(a)
that schematically describes the thermodynamics predicted in the
hadron resonance gas (HRG) model.  The HRG assumes a gas of
non-interacting hadrons in a thermal bath.  Each hadron hardly
contributes to the entire thermodynamics, but the whole sum from
hundreds of hadrons amounts to a substantial portion of the total
pressure.  As a side remark I note that it is a bit subtle whether the
hadron interaction is incorporated or not in the HRG model.  Some
people claim that interaction is correctly taken into account via
higher resonating states.  Such a claim is true if all the $S$-matrix
poles are picked up by existing bound states and/or resonances.  There
may be, however, missing contribution from branch cut associated with
threshold behavior and from hidden states not listed on the particle
database.  The theoretical foundation for the validity of the HRG
estimate should deserve future investigations.

In any case, the fact is that the coincidence between the lattice QCD
and the HRG is becoming better and better up to $\Tc$ as the lattice
QCD approaches the continuum limit and the physical quark mass.
Above $\Tc$, eventually, the HRG badly blows up, which is to be
regarded as the breakdown of the HRG model.  Here, I would point out
two non-trivial problems in this HRG results, which will provide us
with a useful insight when we address dense matter.

The first problem is that we can no longer consider that the rapid
rise in $p/T^4$ should be attributed to color deconfinement.  It could
be induced by hundreds of hadrons even without colored particles at
all.  This is why I wrote; it is not wrong but not completely
satisfactory to say that deconfinement takes place when $p/T^4$ shows
a quick increase.  Then, you may wonder, as a radical extreme, if it
makes logical sense that the system keeps hadrons as relevant degrees
of freedom for any $T$ including $T>\Tc$.  We already know that the
standard HRG is not valid above $\Tc$, but the question is what
exactly breaks down there. Does the hadronic description lose its
meaning at all or is it only a part of assumption made in the HRG that
goes invalid?

This question is deeply related to my second problem; that is, how is
it possible that the thermodynamics is taken over from the HRG to the
QGP as observed in the lattice QCD [see Fig.~\ref{fig:thermo}~(a)]
even though the pressure of the HRG is greater.  It is a firm
thermodynamic principle that any state with a larger pressure would be
more favored.  Therefore, if the HRG is somehow legitimate above
$\Tc$, the hadronic state should be more stable than the QGP, which is
not the case in reality.  It is not very difficult to propose a
consistent picture that resolves my two problems.  Hadrons or mostly
mesons are accompanied by interaction clouds inside of a non-zero
radius $\sim\LQCD^{-1}$.  Thus, the Hagedorn-like behavior of the HRG
pressure should be \textit{saturated} by the interaction clouds of
mesons.  It follows that each border of meson is not really
well-defined and quarks and gluons can hop around between overlapping
wave-functions of blurred mesons.  This is nothing but a percolation
picture of color deconfinement.  So, we do not have to abandon the
hadronic description even above $\Tc$ only if we incorporate not
point-like mesons but extending mesonic wave-functions, and this is
equivalent to introducing colored particles in the thermal system
after all.

Now we have a right answer to the question about the onset of
deconfinement.  Your answer should be that deconfinement takes place
when the growth in $p/T^4$ is saturated due to hadronic interactions.
Such a strongly interacting mesonic state is more naturally and
efficiently handled by (resummed) perturbation theory of hot QCD and
the hard thermal loop approximation is so successful to account for
the lattice thermodynamics from the high temperature side up to the
temperature around $\sim 2\Tc$ as sketched in
Fig.~\ref{fig:thermo}~(b).  We notice that the most non-trivial region
near $\Tc$ is to be characterized by a sort of duality between
strongly interacting mesons and weekly interacting quarks and gluons.
Because this region looks so special, we are tempted to name it, and
actually a well-known name has already been given;  that is, the
strongly-correlation QGP or shortened as the sQGP.\ \  Some other
people named it as the semi-QGP~\cite{Hidaka:2008dr} and this latter
is much better than calling it the sQGP in my opinion.  Here, I would
propose another name;  the Gluesonic Matter in a sense that
\textit{gluonic} degrees freedom $\sim\mathcal{O}(\Nc^2)$ can be
realized in the interacting \textit{mesonic} matter.  You may think
that it is just a matter of terminology, but I would stress that many
people still misunderstand the essential idea of the Quarkyonic
Matter, as we will discuss later, and the alternative name of the
Gluesonic Matter helps us with understanding the Quarkyonic Matter
correctly.

\section{Onset of deconfinement at high baryon density}

We are ready to go into our main discussions on dense QCD matter.
Quark matter was first speculated in neutron
stars~\cite{Itoh:1970uw} simply as a hypothetical state of matter.
You may want to argue that the asymptotic freedom at high density and
the Debye color screening could justify deconfined quarks and gluons
like the QGP at high temperature~\cite{Collins:1974ky}.  The
contemporary QCD theorists are aware, however, that magnetic gluons
cannot be screened and it may be even misleading to identify the
typical scale of interaction with the chemical potential ($\muB$ for
the baryon number or $\muq=\muB/\Nc$ for the quark number).

It is thus a profound and unanswered question how we can retrieve any
information on density-induced deconfinement from QCD itself.
Theoretically speaking, even if the lattice QCD were at work to study
finite-density systems, the Polyakov loop that is an approximate order
parameter for deconfinement at high temperature would be of no use in
order to characterize a state of matter in the low-$T$ and high-$\muB$
region. Then, how can we see deconfinement with any physical
observable?

Let us recall that we have witnessed deconfinement not relying on an
approximate order parameter in the previous section.  That is,
deconfinement from baryonic to quark matter should take place when the
pressure is saturated to be of $\mathcal{O}(\Nc)$ due to baryonic
interactions.  Interestingly enough, the large-$\Nc$ expansion of QCD
rigorously justifies such a physics picture.

\begin{figure}
 \begin{center}
 {\huge (a)} \hspace{0.23\textwidth} {\huge (b)}\\[1em]
 \includegraphics[width=0.15\textwidth]{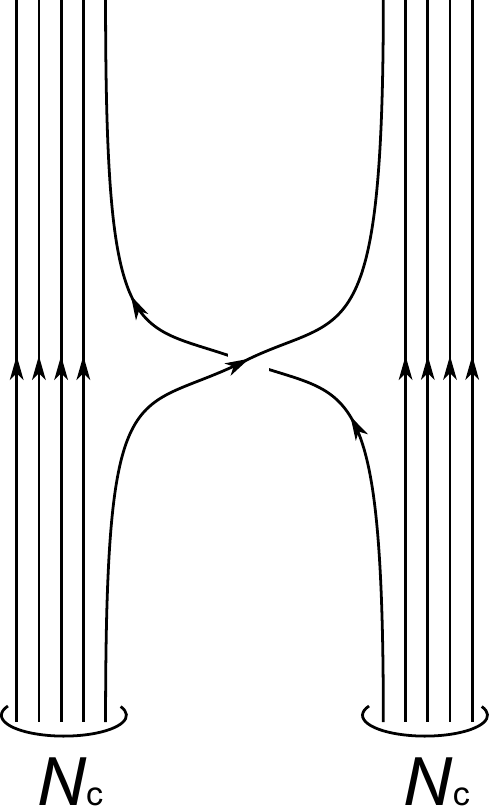} \hspace{6em}
 \includegraphics[width=0.15\textwidth]{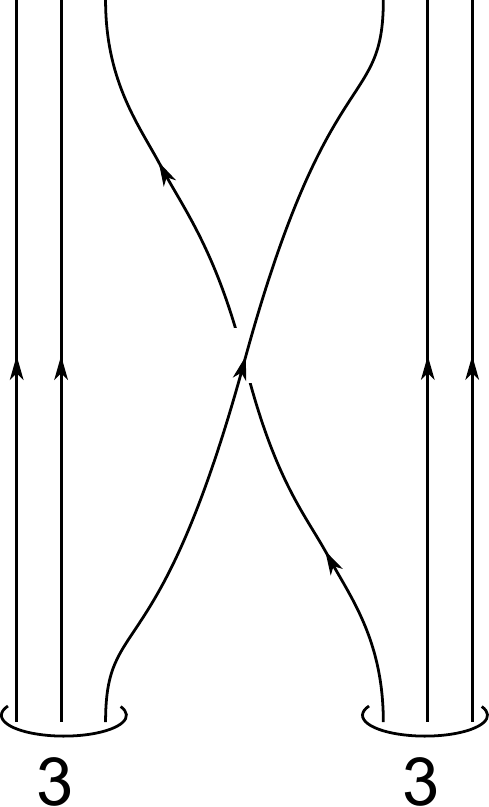}
 \end{center}
 \caption{Schematic figures of the baryon-baryon interaction via quark
   exchange.}
 \label{fig:baryon}
\end{figure}

In the large-$\Nc$ limit baryons interact strongly via quark
exchange.  The constraint that baryons carry no net color allows for
$\Nc$ combinations of quark exchange (or $\Nc^2$ combinations with one
gluon exchange leading to $g^2\sim 1/\Nc$) as seen in
Fig.~\ref{fig:baryon}~(a), which indicates that the baryon interaction
energy is of $\mathcal{O}(\Nc)$.  Now it is obvious that the situation
is quite analogous to what we have seen using the HRG model at
low-$\muB$ and high-$T$.  Therefore, \textit{baryonic} matter is a
dual of \textit{quark} matter as it is, and in this sense, we should
reasonably call it the Quarkyonic Matter~\cite{McLerran:2007qj}.  If
you do not like the name of the Quarkyonic Matter, you can give your
own name such as the ``strongly-correlated quark matter'' etc, but it
is just a matter of terminology.

\begin{figure}
 \begin{center}
 \includegraphics[width=0.4\textwidth]{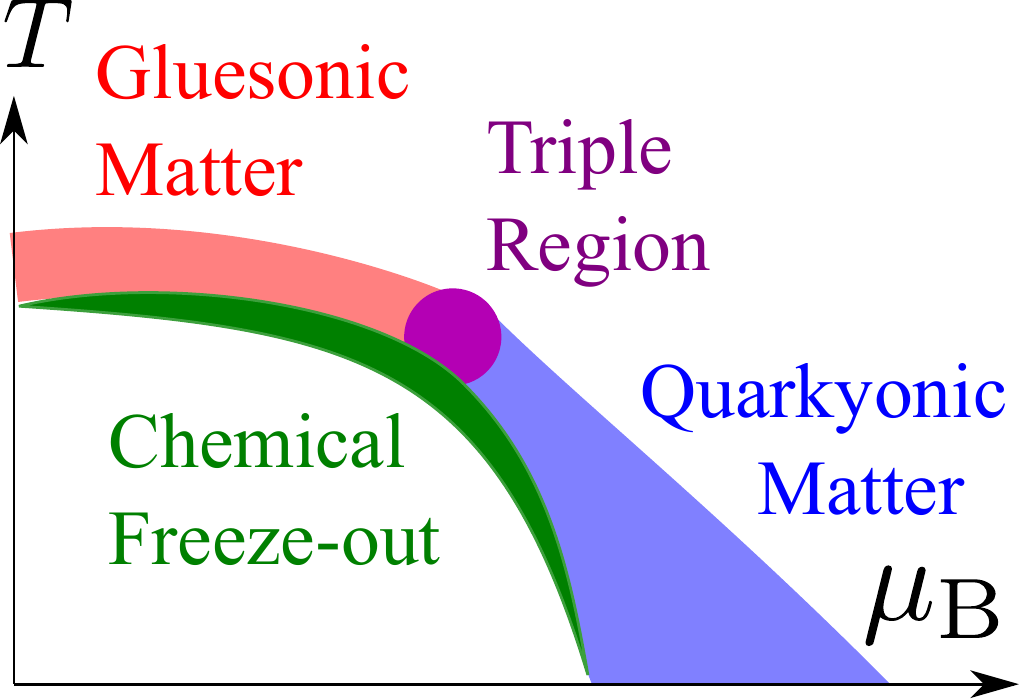}
 \end{center}
 \caption{Schematic figure of the QCD phase diagram separated by the
   Gluesonic Matter in the low-$\muB$ and high-$T$ region, the
   Quarkyonic Matter in the low-$T$ and high-$\muB$ region, and a
   region between them labeled as the Triple Region, which shape a
   boundary along the chemical freeze-out curve in the $\muB$-$T$
   plane.}
 \label{fig:phase}
\end{figure}

Apart from chiral symmetry that is the subject in later discussions,
we can draw a qualitative QCD phase diagram as depicted in
Fig.~\ref{fig:phase}.  The hadronic matter is surrounded by the
chemical freeze-out curve that is associated with the Gluesonic Matter
in the low-$\muB$ and high-$T$ region, the Quarkyonic Matter in the
low-$T$ and high-$\muB$ region, and a transitional region between them
that looks like a triple point or a Triple
Region~\cite{Andronic:2009gj}.

There is one important distinction of the Quarkyonic Matter from the
Gluesonic Matter.  In the large-$\Nc$ limit mesons are non-interacting
particles because the decay constant scales as $f_\pi\sim\sqrt{\Nc}$
and so the interaction vertices should be suppressed accordingly to
make the total amplitude stay finite.  This is why the pressure has a
jump as a function of $T$.  As long as the mesons are excited
dilutely, the interaction clouds are negligible and the pressure
remains to be of $\mathcal{O}(\Nc^0)$.  In contrast to this, in the
Quarkyonic Matter, baryons always interact strongly.  So, it is quite
unlikely that a rapid rise as in Fig.~\ref{fig:thermo}~(a) occurs with
increasing density or chemical potential.  We should therefore
consider that the Quarkyonic Matter extends smoothly from conventional
nuclear matter to quark matter at asymptotically high density (and
this is why some people correctly say that the Quarkyonic Matter is
nothing but nuclear matter, which is not wrong but not completely
satisfactory).  I would emphasize that this is a good news for future
experimental prospects.  We were excited about the so-called sQGP and
why not should we about the counterpart in the high density region
that has an even wider terrain on the phase diagram?  It is not easy
to make any concrete predictions on the properties of the Quarkyonic
Matter, but we should remember that no concrete predictions had
existed before RHIC that would signal for the sQGP or the Gluesonic
Matter.

Coming back to the theoretical consideration, those who are familiar
with dense QCD may have the following question;  where is the color
superconductivity (CSC)?  This is an absolutely decent question.  I
know that some people living in the strict large-$\Nc$ world would
emphasize the irrelevance of CSC too much, making too little of the
reality of CSC.\ \ I believe, however, that we are living in a
world with $\Nc=3$ and we must directly face such a question of how to
reconcile the Quarkyonic Matter and the CSC on equal footing.

\section{Reality of diquarks}

The picture of the Quarkyonic Matter is not necessarily incompatible
with the presence of diquarks.  Since the recognition of the
Quarkyonic Matter originated from the large-$\Nc$ limit where the
diquark interaction is suppressed, you may well think that the
Quarkyonic Matter and the CSC are not cooperative but behave more like
cats and dogs.  Of course, cats and dogs would not always struggle
with each other, and there should be a way how they could live
together peacefully beneath the same roof.

In the large-$\Nc$ world the sources of the baryon density and the
pressure are clearly separated;  the baryon density appears solely
from static baryons and the pressure is dominated by the mesons
between baryons.  Although it is not written explicitly in
Fig.~\ref{fig:baryon}~(a), multiple gluons propagate between exchanged
quarks in the $t$-channel, and the interaction is mediated by not two
quarks but a meson.  So, our intuition based on a Fermi gas with Fermi
surface completely breaks down.  From the diagrammatic point of view,
however, there is no reason why we should exclude a resonating state
of two quarks in the $s$-channel.  Such an intermediate state of two
quarks is not a color singlet, but as long as it emerges through a
virtual state, there is no practical problem.  Then, with ladder-type
resummation of gluonic processes, Fig.~\ref{fig:baryon}~(a) actually
describes:
\begin{itemize}
 \item Two baryons having a strong interaction mediated by meson
   exchange in the $t$-channel.
 \item Two baryons having a mixture with two $(\Nc-1)$-quark objects
   and a diquark in the $s$-channel.
\end{itemize} 
Here you may concern that the four-quark interaction in the diquark
channel should be suppressed by $1/\Nc$ than that in the meson
channel, but this $1/\Nc$ is compensated for by $\Nc$-colored
diquarks.  We see that the latter picture of a diquark mixture is more
understandable on the intuitive level.  In the $\Nc=3$ world, as
illustrated in Fig.~\ref{fig:baryon}~(b), the interpretation is even
more intuitively appealing.  The $(\Nc-1)$-quark part is also the
diquark and a quark-diquark mixture represents the interacting
intermediate state.  Hence, we shall adopt a working definition of the
Quarkyonic Matter characterized by a mixture of diquarks.

It is an empirically established idea to construct the baryon
wave-function as a bound state of a diquark and a quark.  In the quark
model such a construction simply refers to the group theoretical
structure of color indices, and it does not necessarily require the
reality of diquarks.  In the Faddeev equation in the four-quark
interacting model the diquarks acquire more reality along the line of
Fig.~\ref{fig:baryon}~(b).  Besides, the famous $\Delta I=1/2$ rule of
non-leptonic weak decays suggests the presence of spatially compact
diquark inside of the baryon wave-function.

Once we admit the presence of strong diquark correlation, we have an
immediate problem.  If we can define the diquark mass (or precisely
speaking, if the diquark spectral function has a prominent peak at
some frequency), and if $\muB$ exceeds roughly $\Nc/2$ times the
diquark mass, we cannot avoid the Bose-Einstein condensation of
diquarks.  This means, if the diquark mass is not so far from $2/\Nc$
times the baryon mass or twice of the constituent quark mass, the CSC
is unavoidable even in the vicinity of nuclear matter.  At a first
glance you may be inclined to think that the CSC near nuclear matter
is an artifact of the naive diquark model.  Before making any
judgment, we should think twice about the reality of the CSC in
nuclear matter, however.

It is very unlikely that nuclear matter exhibits a sharp first-order
phase transition to quark matter, though it was a conventional
approach to postulate the equation of state of dense
matter~\cite{Baym:1976yu}.  If we introduce the diquark degrees of
freedom between nuclear and quark matter, everything seems to be quite
consistent.  As we already discussed, deconfinement cannot be usually
guaranteed by the screening phenomenon in quark matter, but if the
CSC (or the color-flavor-locked state, strictly speaking) occurs, all
gluons are gapped as a result of the Meissner effect.  Then, the
perturbation theory becomes well-defined and the magnetic sector leads
to the celebrated enhancement of the gap energy.  In this way, it is
not such a surprising proposition to \textit{define} deconfinement of
dense matter by the formation of the diquark condensate.  Then, a CSC
expert would pose the following question;  how can you define the CSC
or the diquark condensate in a gauge invariant way?  This is not an
academic question but rather a very pragmatic question.  On the
academic level the simplest answer is that I fix a gauge so that the
diquark condensate can take a non-zero value.  In real experiment,
however, the physical observable should be gauge invariant and the
diquark condensate is not detectable in principle.  This
undetectability is nicely summarized in terms of the Quark-Hadron
Continuity~\cite{Schafer:1998ef,Alford:1999pa}.  From this theoretical
point of view of the Quark-Hadron Continuity, in fact, there is no
contradiction even if we assume a small fraction of the diquark
condensate inside of ordinary nuclear matter.  In nuclear theory one
of the most well-known calculation schemes is the
Hartree-Fock-Bogoliubov theory and the pairing interaction leads to a
finite pairing gap.  The diquark condensate would break the U(1)
symmetry associated with the baryon number conservation as well as
chiral symmetry.  Because we know that both the U(1) symmetry and
chiral symmetry are broken in superfluid nuclear matter, no symmetry
argument prohibits the existence of the diquark condensate.  In other
words, if the existence is not ruled out by the symmetry reason, we
should think that it must be there.

The reality of diquarks in nuclear matter would open an intriguing
opportunity for upcoming experimental attempts to pursue compressed
baryonic matter.  Diquarks are fundamental building blocks of exotic
hadrons composed from more constituent quarks than $q\bar{q}$ or
$qqq$.  There is an idea to manifest the $qq$ part only by inserting a
heavy flavor $Q$ into a state; $Qqq$, which is actively promoted by
Japanese hadron physicists (learned from private communications with
A.~Hosaka and M.~Oka).  The high baryon density is another way to
manifest the diquark correlation.  Here, it must be mentioned that the
diquark correlation is usually seen in momentum space, as in the
Cooper pair of electrons in the ordinary BCS theory.  Hence, whenever
we talk about the experimental challenge to see the diquark
correlation, it should be clearly stated which of the correlation in
momentum space and in configuration space is sensitive to the proposed
observable.  We should always keep in mind that demanding the presence
of spatially compact diquark is a very strong assumption and there is
no theoretical argument that can endorse any strong diquark
correlation in space, that is the case also when the CSC is turned on.
Nevertheless, the diquark should be enumerated as the top-priority
keyword for the expected paradigm shift from the Gluesonic Matter to
the Quarkyonic Matter in the near future.

\section{Topological effects coupled to the baryon density}

Let us change the subject from the state of matter to a fancy
phenomenon with the quantum anomaly.  In condensed matter physics the
Magnetoelectric Effect has been long known;  if you impose $\bE$ on a
special material, you observe $\bB$ in parallel to $\bE$, which
implies a source term, $\theta\bE\cdot\bB$, in the effective
Lagrangian.  Such a term is called the $\theta$-term and the
topological insulator corresponds to the case with $\theta=\pi$, for
example.  In the nucleus-nucleus collision, a very strong magnetic
field whose energy scale is comparable with $\LQCD$ is expected if the
impact parameter is non-zero.  Then, if the QGP accommodates the
$\theta$-term, it is quite conceivable to anticipate some topological
effect in analogy to the Magnetoelectric Effect.

The most well-known example of the topological effect along this line
is the Chiral Magnetic Effect~\cite{Fukushima:2012vr}.  The severest
problem in the Chiral Magnetic Effect is that it is sensitive to the
Local Parity Violation and thus the net effect averaged over space
and/or collision events should be vanishing.  So, what we can see
experimentally is only the fluctuation that is parity even, and in
principle, the signal is not separable from the background.  Even
though the theoretical and experimental efforts are continuing for the
Chiral Magnetic Effect, we further need a breakthrough to invent
some better observable than the fluctuation of the charge separation;
otherwise, we can never conclude anything for or against the Chiral
Magnetic Effect.

The situation is significantly improved as soon as the density comes
into the game of the quantum anomaly.  The reason for this is obvious
for theorists;  the chemical potential is the zeroth component of the
gauge field, and so, the finite-density physics is always accompanied
by the gauge dynamics.  A typical example is found in the density
origin in (1+1)-dimensional gauge theories, in which a finite density
arises purely from the Wess-Zumino term.  Then, what is the
theoretical prediction if we have a finite $\muB$ and a strong $\bB$?
The answer is recently referred to as the Chiral Separation Effect and
the axial current $\bj_5 \propto \muB \bB$ is generated.  You may
think that $\bj_5$ in the Chiral Separation Effect would induce the
chirality imbalance in the same way as the charge separation in the
Chiral Magnetic Effect, but the chirality is not a conserved charge.
It decays through a finite mass and a topologically winding
configuration.  Still, the Chiral Separation Effect at finite density
is much more advantageous than the Chiral Magnetic Effect because it
is a net effect and the spatial and/or event average does not wash it
out.  As we stated, chirality is not a conserved charge, but the
helicity is a good quantum number, and so the chirality decay may lead
to an interesting consequence about, so to speak, helicity
transmutation.  Such a possibility has been addressed in the context
of neutron star physics~\cite{Ohnishi:2014uea}, and in my belief, this
idea could be usefully imported to the analysis of lower energy
heavy-ion collisions.

The physical interpretation of the topological currents has a subtle
aspect.  In theory the current just means an expectation value of
spinor bilinear operator in a certain channel.  If we have a
condensate of the axial vector meson for some reason, we also acquire
a non-zero $\bj_5$, and is this really a current that flows in real
time?  It is not so easy to let our imagination work, but the answer
is yes.  The vector condensation and the supercurrent are usually
identifiable.  Then, when we consider the ground state of dense QCD
matter, the condensation of $\bj_5$ or its mean-field effect must be
taken into account just like the mean-field density effect from the
vector interaction.  To this end, we have to move to our final topic;
the possibility of the inhomogeneous chiral condensates.

\section{Inhomogeneous chiral condensates}

It is an old but still vital idea that the spatial modulation occurs
at high density, so that the genuine ground state should be
characterized by inhomogeneous chiral condensates.  A nice review by
pioneering researchers, Buballa and Carignano, is quite recommendable
for interested readers~\cite{Buballa:2014tba}.  The idea is traced
back to the $p$-wave pion condensate in nuclear matter.  Although the
pion condensate is disfavored by the Gamow-Teller giant resonance, it
is not yet completely excluded, and moreover, the idea is still alive
as a realistic possibility inside of quark matter.

\begin{figure}
 \begin{center}
 \includegraphics[width=0.28\textwidth]{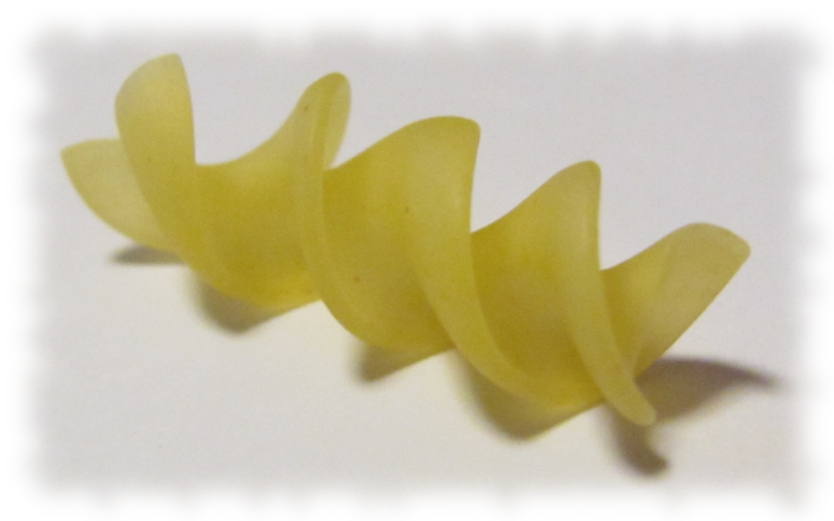}
 \end{center}
 \caption{Schematic view of the chiral spiral in QCD; you can easily
   find this at kitchen as I did.}
 \label{fig:escher}
\end{figure}

The simplest way to introduce spatial inhomogeneity is to postulate a
one-dimensional spiral structure in the scalar and the pseudo-scalar
channels as:
$\sigma\sim \cos(q z)$ and $\pi^0\sim \sin(q z)$ along the $z$
direction.  Such a system is not much different from the homogeneous
case, and in fact, if the density is calculated, it turns out to be a
spatial constant.  This type of chiral condensate is called the dual
chiral density wave~\cite{Nakano:2004cd} or the chiral spiral.  In the
space of $\sigma$ and $\pi^0$ condensates, the chiral spiral looks
like pasta called fusilli as in Fig.~\ref{fig:escher}~\footnote{In the
  talk at QM2015 I showed a famous painting by Escher, ``Spirals'',
  which is a good analogy to the chiral spiral of QCD, and this
  beautiful painting was supposed to be presented here.  The
  M.C.\ Escher Company, however, charged 75 euros for copyright-fee,
  35 euros for handling, and 75 euros additionally per electronic
  image.  So, I gave up demonstrating artistic nature of physics and
  replaced it with another branch of human culture; gastronomic
  creation of Italy.}.  I would say that the QCD ground state could be
as fascinating as pasta al dente.

Here we are not going into details, but let us mention on the ``facts''
only.  Some model calculations predict the so-called QCD critical
point, and some others predict a smooth crossover on the whole QCD
phase diagram without the critical point.  This has been a status of
the QCD critical point search, and unlike this frustrating situation,
all the model calculations predict spatially inhomogeneous chiral
condensates in the high-density region, which approximately overlays
the region of the Quarkyonic Matter in Fig.~\ref{fig:phase}.  Such
coincidence of the inhomogeneous region and the Quarkyonic Matter is
not accidental;  the chiral condensate in the large-$\Nc$ world should
be inhomogeneous.  This can be understood from the fact that baryons
are infinitely heavy in the large-$\Nc$ limit.  The optimal
configuration of baryonic matter is thus a crystal of static baryons
and then the chiral condensate becomes smaller locally near baryons
than the vacuum value.

A frequently asked question about the inhomogeneous chiral condensates
is that it may be unstable against mesonic fluctuations.  This is a
very reasonable question.  In condensed matter physics, a hypothetical
crystalline superconducting state (called the FFLO state) has been a
long-standing issue.  In most cases, such an exotic state has the
lowest energy in the mean-field analysis, while it goes away once
fluctuations of Nambu-Goldstone bosons or spatial rotations are
considered.  In QCD, therefore, we should pay a serious attention to
the roles played by pion fluctuations and rotations of a finite-sized
fireball to test the stability of the inhomogeneous chiral
condensates.  There are some works on this matter, but the final
sentence still awaits to be announced.

Naturally, such a spiral would be affected by the topological current
$\bj_5$ if $\bj_5\neq0$, since the chiral spiral represents a flow of
chirality as perceived from Fig.~\ref{fig:escher}.  Therefore, when we
talk about the chance to study baryonic matter in the heavy-ion
collision, we should deal with both spatial modulations and
topological currents.  There are not many theoretical efforts yet
in this direction, probably because of uncontrollable ambiguity in
model treatments.  So far, there is one concrete result in the
large-$\Nc$ limit using the holographic QCD
model~\cite{Fukushima:2013zga}, which suggests that the spatial
modulation becomes less favored with stronger axial-vector
interaction, $\bj_5\cdot\bj_5$, that should be significantly enhanced
by the topological current $\bj_5\neq0$.  This conclusion is
consistent with the observation of the $p$-wave pion condensate
diminished by the Landau-Migdal interaction in the spin-isospin
channel.

If the inhomogeneous chiral condensates can survive in reality, the
experimental survey for this structure should go in a similar fashion
to the Local Parity Violation.  Instead of parity-odd domains, we
should search for dense bubbles.  The problem is again that we cannot
make a discovery in a qualitative sense but the analysis always relies
on quantitative comparisons, and such a strategy does not work unless
we establish undoubted theoretical predictions.  This may sound like a
difficult task, but I am rather optimistic.  Before the age of RHIC
and LHC, who could have foreseen such a big success of the thermal fit
and the HRG model?  We certainly need some refinement of those
descriptions, probably with several mean-fields as in the relativistic
mean-field model of nuclear matter or with new degrees of freedom like
diquarks.  With sufficient data of hadron multiplicity and their
fluctuations at lower energies, it is simply a matter of time to come
by a reliable baseline in order to diagnose the intrinsic properties
of dense QCD matter.  I would not guarantee anything particularly
interesting between cold QGP and hot nuclear matter, that is the
regime accessible by future heavy-ion programs, but I can at least say
that this experimentally accessible regime could definitely provide us
with lots of hints to physics questions that nobody can answer at
present.

Lastly, I would like to express my thanks to G.~Baym, M.~Buballa,
T.~Hatsuda, A.~Hosaka, L.~McLerran, M.~Oka, J.~Pawlowski,
M.~Stephanov, T.~Tatsumi, N.~Yamamoto.  They often gave me a hard time
with unanswerable questions, sometimes opened my eyes to a new idea,
and occasionally shared physics motivation with me.  The contents of
this contribution are largely influenced by stimulating discussions
with these brilliant researchers.



%% References


\bibliographystyle{elsarticle-num}
\bibliography{fuku.bib}



%\begin{thebibliography}{00}


%\bibitem{ref1} J. van der Geer, J.A.J. Hanraads, R.A. Lupton, J. Sci. Commun. 163 (2000) 51-59. 
%\bibitem{ref2} W. Strunk Jr., E.B. White, The Elements of Style, third ed., Macmillan, New York, 1979. 

%\end{thebibliography}

\end{document}